\documentclass[fleqn,twoside]{article}
\usepackage{espcrc2}

\usepackage{graphicx}
\usepackage{axodraw}
\usepackage[figuresright]{rotating}

\newcommand{\AmS}{{\protect\the\textfont2
  A\kern-.1667em\lower.5ex\hbox{M}\kern-.125emS}}
\newcommand{\be}{\begin{equation}}
\newcommand{\ee}{\end{equation}}

\title{Two-Loop QED Heavy-Flavor Contribution to Bhabha Scattering%
        \thanks{{\tt Work supported by the Swiss National Science Foundation
    (SNF) under contract 200020-117602. Z\"urich preprint number ZU-TH 11/08.}}
}

\author{R. Bonciani\address[MCSD]{Institut f\"ur Theoretische Physik,
        Universit\"at Z\"urich, CH-8057, Z\"urich, Switzerland }
     and
       A. Ferroglia \addressmark
}
\begin{document}

\begin{abstract}
We briefly review the status of the calculation of
next-to-next-to-leading order corrections  to large angle Bhabha
scattering in pure QED. In particular, we focus on the analytic
calculation of the two-loop virtual corrections involving a
heavy-flavor fermion loop, which  was recently completed. We
conclude by assessing the numerical impact of these corrections on
the Bhabha scattering cross section at colliders operating at a
center of mass energy of 1 GeV and at the future ILC. \vspace{1pc}
\end{abstract}

\maketitle

\section{Introduction}

Bhabha scattering \cite{Bha} is the scattering process of an
electron and a positron going into an electron-positron final
state. The scattering of electrically changed particles is always
accompanied by electromagnetic radiation, so that it becomes
necessary to consider Bhabha scattering events of the type $e^+ \,
e^- \to e^+ \, e^- + n \gamma$.

Since Bhabha scattering has a large cross section and  involves
charged leptons in the final state, it is a process that can be
measured experimentally with very high  accuracy. For example, in
the KLOE/DA$\Phi$NE experiment with a center of mass energy of
$\sqrt{s} \sim 1$ GeV, the cross section integrated over the
allowed phase space region with scattering angle $55^\circ <
\theta < 125^\circ$ is of about 400 nb, known with 1 permille
accuracy.

Both at high energy $e^+-e^-$ colliders  and at $e^+-e^-$
colliders operating at a intermediate center of mass energies
($\sqrt{s} \sim 1-10$  GeV), it is possible to find kinematic
regions in which the Bhabha scattering cross section is dominated
by QED. This is the reason why the Bhabha scattering cross section
can be calculated with high precision in perturbation theory,
including next-to-next-to-leading order (NNLO) corrections.

Due to the nature of the two properties listed above, the Bhabha
scattering  was chosen for the luminosity determination at LEP, at
the flavor factories  (DA$\Phi$NE, VEPP-2M, etc.), and it will be
employed for the luminosity measurement at the future
International Linear Collider (ILC).
At colliders operating at a center of mass energy $\sqrt{s} \sim
1-10$ GeV, the luminosity  is measured by observing Bhabha events
at large scattering angles, $\theta \sim 90^{\circ}$; in the past,
at colliders operating at high center of mass energy such as LEP
($\sqrt{s} \sim 90,200$ GeV), the luminosity was  measured by
observing Bhabha events at small scattering angles, $\theta \sim
1^{\circ}$. At the ILC ($\sqrt{s} \sim 500$ GeV), in order to
disentangle the luminosity spectrum, it will be necessary to
control both regions, $\theta \sim 1^{\circ}$ and $\theta \sim
90^{\circ}$.

The luminosity ${\mathcal L}$ is defined as the ratio between the
rate of events of a reference scattering process divided by the
theoretical cross section of  the same process: ${\mathcal L} =
N/\sigma_{th}$. It enters, as a normalization factor, in the
measurements of all other cross sections. Therefore, it has to be
known with a very high accuracy. Since $\delta L = \delta N +
\delta \sigma_{th}$, and $\delta N$ is to a large extent under
control, a crucial  point is to reduce  the theoretical
uncertainty $\delta \sigma_{th}$ to the greatest extent. This
involves several steps, that we list schematically below.
\begin{itemize}
\item[i)] To begin, one needs a precise knowledge of the matrix elements,
that is, to control  the perturbative corrections to the basic
process $e^+ \, e^- \to e^+ \, e^- + n \gamma$. At the level of
accuracy required by the experimental measurement, the NNLO
corrections have to be taken into account. Therefore, from a
diagrammatic point of view, it is necessary to evaluate two-loop
$2 \to 2$ Feynman diagrams as well as one-loop $2 \to 3$ and
tree-level $2 \to 4$.
\item[ii)] Secondly, the theoretical cross section has to be
inclusive of the soft-photon radiation and hard-collinear photon
radiation.
\item[iii)] Finally, it is necessary to consider the specific experimental set up.
Usually detectors do not cover the whole solid angle but just a
portion of it; when calculating the cross section, it is necessary
to account for these experiment specific geometric constraints.
\end{itemize}
An accurate theoretical description of the Bhabha scattering must
consider all these aspects. For this purpose, the best tools are
Monte Carlo event generators (MC) that merge the fixed-order
calculations with parton showers that take into account collinear
and soft electromagnetic radiation. In the past, many MCs for the
evaluation of the Bhabha scattering cross section were developed
\cite{BHAGENF,Cacciari:1995fq,BHLUMI,BHWIDE,MCGPJ,BABAYAGA,BABAYAGA-1}.
For a detailed discussion of their properties see \cite{Montagna}.

In the rest of this proceeding, we will concentrate on the analytic calculation
of the perturbative corrections to the matrix element.

\section{Perturbative Corrections}

Since the theoretical error on the Bhabha scattering differential
cross section directly affects the precision of the luminosity
determination, in recent years a significant effort was devoted to
the calculation of  perturbative corrections to this scattering
process. The NLO corrections are well known in the full Standard
Model \cite{Bhabha1loop}.  For what concerns the electroweak
corrections at NNLO, only the logarithmic enhanced terms are known
\cite{KA}. In pure QED,  the situation is significantly different.
The first complete diagrammatic calculation of the two-loop QED
virtual corrections to Bhabha scattering can be found in
\cite{Bern}. However, this result was obtained by setting the
electron mass $m_e$ to zero from the start, and by employing
dimensional regularization (DR) to regulate both soft and
collinear divergencies.

Today, the complete set of NNLO corrections to Bhabha scattering
in pure QED have been evaluated using $m_e$ as the  collinear
regulator, as required in order to include these fixed order
calculations in MCs.
The two-loop
Feynman diagrams involved in the calculation can be divided in three gauge
independent sets: {\em i)} diagrams without fermion loops (``photonic''),
{\em ii)} diagrams involving a closed electron loop, and {\em iii)}
diagrams involving a closed loop of a fermion different from the electron.
%

A large part of the NNLO ``photonic'' corrections can be obtained
in a closed analytic form, including the full dependence on the
electron mass, using a technique that has by now become standard
in multi-loop calculations. This technique is based upon the
Laporta algorithm \cite{Laportaalgorithm} for the reduction of the
Feynman diagrams to  Master Integrals (MIs), and further based
upon the differential equation method \cite{DiffEq} for the
analytical evaluation of the latter. With this technique, it is
possible to calculate the NNLO corrections to the form factors
\cite{Bonciani:2003te} (see also the analogous cases for heavy
quarks \cite{Bernreuther:2004ih} and Higgs \cite{Aglietti:2004nj})
and to provide the photonic corrections to the cross section with
the exception of the ones originating from two-loop boxes
\cite{BF}.
The calculation of the two-loop photonic boxes retaining the full dependence
on $m_e$ is beyond the reach of this technique (partial
results can be found in \cite{Smirnov}). However, the full dependence on $m_e$
is not phenomenologically needed \cite{BF}.
Fortunately, the physical problem exhibits a well defined mass
hierarchy. The mass of the electron is always very small compared
with  other kinematic invariants, and it can be safely neglected
everywhere except when it acts as a collinear regulator.
The collinear structure of the NNLO corrections is the
following:
\be
\frac{d \sigma^{(NNLO)}}{d \sigma^{(\mbox{\tiny{Born}})}}  =
\frac{\alpha^2}{\pi^2} \sum_{i=0}^{2} \delta^{(i)} L_e^i +
\mathcal{O}\left(\frac{m_e^2}{s}, \frac{m_e^2}{t}\right) \, ,
\label{exp-photonic}
\ee
where $L_e=\ln{(s/m_e^2)}$ and where $\delta^{(i)}$ are
functions of $\theta$ and of the mass of the
heavy fermions involved in the virtual corrections. The approximation given
by Eq.~(\ref{exp-photonic}) is sufficient for a phenomenological description of
the process.

In the case of photonic corrections, the coefficients of the
square and single collinear logarithm in Eq.~(\ref{exp-photonic})
were obtained in \cite{L2,Bas}. The precision required for
luminosity measurements at $e^+ e^-$ colliders demanded the
calculation of the non-logarithmic coefficient, that was  obtained
in \cite{Pen} through the {\em infrared matching} to the massless
approximation. The technique of \cite{Pen} allowed the
reconstruction of the photonic differential cross section in the
$s \gg m_e^2 \neq 0$ limit from the calculation in \cite{Bern},
where $m_e$ was set to zero from the  start. The method employed
in \cite{Pen} involves a change of regularization scheme for the
collinear divergencies originating from a vanishing electron mass.
A method based on a similar principle was subsequently  developed
in \cite{Mitov:2006xs,BecMel}; the authors of \cite{BecMel}
confirmed the result of \cite{Pen}.


The corrections involving a closed electron loop were calculated
diagrammatically with the Laporta algorithm and the differential
equation method in \cite{electronloop}. The analytic result, which
retains the full dependence on $m_e$, was  expressed in terms of
one- and two-dimensional Harmonic Polylogarithms (HPLs) of maximum
weight three \cite{HPLs}.
By expanding the cross section in the limit $s, |t| \gg m_e^2$,
the ratio of the NNLO corrections to the Born cross section can be
written as in Eq.~(\ref{exp-photonic}). However, the series starts
with a cubic instead of a square collinear logarithm. As expected,
the cubic collinear logarithm cancels out once the contribution of
the soft pair production graphs is added to the virtual and soft
corrections \cite{Arbuzov:1995vj}.
The analytic expressions of the expansion of the results of
\cite{electronloop} was confirmed in \cite{Act,BecMel}.

\section{Heavy Flavor Corrections}

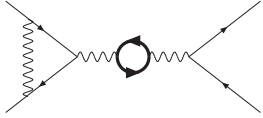
\begin{figure}
\begin{center}
\vspace*{3mm}
\begin{picture}(0,0)(0,0)
\SetScale{0.6}
\SetWidth{0.5}
\ArrowLine(80,-35)(35,0)
\ArrowLine(35,0)(80,35)
\Photon(-66,-25)(-66,23.5){3}{8}
\ArrowLine(-35,0)(-80,-35)
\ArrowLine(-80,35)(-35,0)
\Photon(10,0)(35,0){3}{3.5}
\Photon(-35,0)(-10,0){3}{3.5}
\SetWidth{2}
\ArrowArc(0,0)(10,0,180)
\ArrowArc(0,0)(10,180,0)
\end{picture}
\end{center}
\caption{\label{fig2} \small Reducible two-loop vertex diagram contributing to
the heavy flavor corrections. These kind of diagrams contain single electron
mass logarithms.}
\end{figure}

\begin{figure}
\vspace*{-5mm}
\[
\hspace*{1.1cm}
\vcenter{
\hbox{
  \begin{picture}(0,0)(0,0)
\SetScale{.6}
  \SetWidth{.5}
\ArrowLine(-50,25)(-25,25)
\ArrowLine(-25,25)(25,25)
\ArrowLine(25,25)(50,25)

\Photon(-25,25)(-25,-25){3}{8}
\Photon(25,25)(25,-25){3}{8}

\ArrowLine(-25,-25)(-50,-25)
\ArrowLine(25,-25)(-25,-25)
\ArrowLine(50,-25)(25,-25)
\end{picture}}}
\hspace*{1.2cm}
+
\hspace*{1cm}
\vcenter{
\hbox{
  \begin{picture}(0,0)(0,0)
\SetScale{.6}
  \SetWidth{.5}
\ArrowLine(-50,25)(-25,25)
\ArrowLine(-25,25)(25,25)
\ArrowLine(25,25)(50,25)

\Photon(-25,25)(25,-25){3}{10}
\Photon(25,25)(-25,-25){3}{10}

\ArrowLine(-25,-25)(-50,-25)
\ArrowLine(25,-25)(-25,-25)
\ArrowLine(50,-25)(25,-25)
\end{picture}}
}
\hspace*{1.1cm}
 = \hspace*{0.1cm} \mbox{\begin{tabular}{c}Free of \\ collinear  \\ poles\end{tabular}}
\]
\vspace*{3mm}
\[
\hspace*{1.1cm}
\vcenter{
\hbox{
  \begin{picture}(0,0)(0,0)
\SetScale{.6}
  \SetWidth{.5}
\ArrowLine(-50,25)(-25,25)
\ArrowLine(-25,25)(25,25)
\ArrowLine(25,25)(50,25)
\Photon(-25,25)(-25,-25){3}{8}
\Photon(25,25)(25,10){3}{3}
\Photon(25,-25)(25,-10){3}{3}
\ArrowLine(-25,-25)(-50,-25)
\ArrowLine(25,-25)(-25,-25)
\ArrowLine(50,-25)(25,-25)
  \SetWidth{2}
\ArrowArc(25,0)(10,-90,-270)
\ArrowArc(25,0)(10,-270,-90)
\end{picture}}}
\hspace*{1.2cm}
+
\hspace*{1cm}
\vcenter{
\hbox{
  \begin{picture}(0,0)(0,0)
\SetScale{.6}
  \SetWidth{.5}
\ArrowLine(-50,25)(-25,25)
\ArrowLine(-25,25)(25,25)
\ArrowLine(25,25)(50,25)
\Photon(-25,25)(25,-25){3}{10}
\Photon(25,25)(7,7){3}{4}
\Photon(-25,-25)(-7,-7){3}{4}
\ArrowLine(-25,-25)(-50,-25)
\ArrowLine(25,-25)(-25,-25)
\ArrowLine(50,-25)(25,-25)
  \SetWidth{2}
\ArrowArc(0,0)(10,-90,-270)
\ArrowArc(0,0)(10,-270,-90)
\end{picture}}
}
\hspace*{1.1cm}
 = \hspace*{0.1cm} \mbox{\begin{tabular}{c}Free of \\ collinear  \\ poles\end{tabular}}
\]
\vspace*{-9mm} \caption{Cancellation of the collinear poles among
one- and two-loop box diagrams in the Feynman gauge. The diagrams are
calculated by setting $m_e = 0$ from the start.} \label{fig3}
\vspace*{-4mm}
\end{figure}
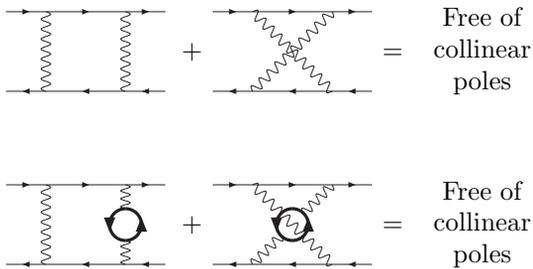

We consider now the corrections originating from two-loop Feynman
diagrams with a heavy fermion loop.
Since this set of corrections involves an additional mass scale
with respect to the corrections analyzed in the previous section,
a direct diagrammatic calculation is in principle a more
challenging task.
Recently, in \cite{BecMel} the heavy flavor NNLO corrections to
the Bhabha scattering cross section were obtained in the limit $s,
|t|, |u| \gg m_f^2 \gg m_e^2$, where $m_f^2$ is the mass of the
heavy fermion in the loop. This result was immediately confirmed
in \cite{Act}.
However, the results obtained in this approximation cannot be
applied to the case in which the $\sqrt{s} < m_f$ (such as in the
case of  tau loops at $\sqrt{s} \sim 1$ GeV), and they apply only
to a relatively narrow angular region perpendicular to the beam
direction when $\sqrt{s}$ is not very much larger than $m_f$ (such
as in the case of top quark loops at ILC). It was therefore
necessary to calculate the heavy flavor corrections to Bhabha
scattering assuming only that $s, |t|, |u|, m_f^2 \gg m_e^2$.
The technical problem is simplified by carefully considering  the
structure of the collinear singularities of this set of
corrections. The ratio of the NNLO heavy flavor corrections to the
Born cross section is given by
\be \label{exp-heavyflavor}
\frac{d \sigma^{(2,\mbox{\tiny{HF}})}}{d \sigma^{(\mbox{\tiny{Born}})}}  =
\frac{\alpha^2}{\pi^2} \sum_{i=0}^{1} \delta^{(\mbox{\tiny{HF},i})} L_e^i +
\mathcal{O}\left(\frac{m_e^2}{s}, \frac{m_e^2}{t}\right) \, .
\ee
It is possible to prove that, in a physical gauge, all the
collinear singularities factorize and can be absorbed in the
external field renormalization \cite{FreTay}. This observation has
two important consequences in the case at hand. The first
consequence is that box diagrams are free of collinear
divergencies in a physical gauge; since the sum of all boxes forms
a gauge independent block, it can be concluded that  the sum of
all box diagrams is free of collinear divergencies in any gauge.
The second consequence is that the single collinear  logarithm in
Eq.~(\ref{exp-heavyflavor}) arises from vertex corrections only.
Moreover, if one chooses on-shell UV renormalization conditions,
the irreducible two loop vertex graphs are free of collinear
singularities.
Therefore, among all the two-loop diagrams contributing to the
NNLO heavy flavor corrections to Bhabha scattering, only the
reducible vertex corrections of  Fig.~\ref{fig2} are
logarithmically divergent in the $m_e \to 0$
limit\footnote{Additional collinear logarithms arise also from the
interference of one-loop vertex and one-loop self-energy
diagrams.}, and they can be easily calculated even though they
involve two different mass scales.
By taking advantage of these two facts we were recently able to
obtain the NNLO heavy flavor corrections to the Bhabha scattering
differential cross section \cite{hfbha}, assuming only that $s,
|t|, |u|, m_f^2 \gg m_e^2$. In particular, in obtaining the
analytic expression for the NNLO cross section, we worked in the
Feynman gauge, setting $m_e =0$ from the start in all the diagrams
with the exception of the reducible ones of the kind shown in
Fig.~\ref{fig2} and in the interference of one-loop graphs. This
procedure allowed us to effectively eliminate a mass scale from
the two-loop boxes. The latter could then be evaluated with the
techniques already employed in the diagrammatic calculation of the
electron loop corrections\footnote{The necessary MIs can be found
in \cite{hfbha,Kalm,MIsHeavy}}.
In this approach, individual box diagrams  are singular in the
$m_e \to 0$ limit (since they are calculated in the Feynman
gauge); their collinear singularities appear as additional poles
in the dimensional regulator $\epsilon$. However, it is easy to
prove that such divergencies cancel in the sum of all the box
diagrams, as schematically illustrated at the one- and two-loop
level in Fig.~\ref{fig3}.
%
%
By expanding the analytic results of \cite{hfbha}, it was possible
to check the heavy flavor cross section in the $s,|t|,|u| \gg
m_f^2 \gg m_e^2$ limit, which was previously known
\cite{BecMel,Act}.
At intermediate energy colliders such as DA$\Phi$NE, the exact
dependence on $m_f$ of the results of \cite{hfbha} allows us to
account for the contribution of muons, taus, $b$- and $c$-quark
loops to the Bhabha scattering cross section.
At high-energy colliders such as the ILC,  the top quark
contribution can also be exactly evaluated.
In the case in which the heavy flavor fermion is a quark, it was
simple to modify the calculation of the two-loop self-energy
diagrams to obtain the mixed QED-QCD corrections \cite{hfbha}.
An alternative numerical approach to the calculation of the heavy
flavor corrections to Bhabha scattering, based on dispersion
relations, was pursued in \cite{Actis:2007}. With this method it
is also possible to evaluate the contribution of the light quarks
vacuum polarization to the Bhabha scattering cross section, that
can be obtained by convoluting the kernel functions with the data
concerning the cross section of the process $e^+ e^- \to$ hadrons.

\section{Conclusions}

The numerical impact of the photonic and electron loop QED
corrections to the Bhabha scattering cross section at flavor
factories was carefully examined in \cite{BABAYAGA-1,Montagna}, in
the context of the MC BABAYAGA. A similar analysis of the heavy
flavor NNLO corrections is not yet  available. However, it is
possible to evaluate numerically  the NNLO heavy flavor
corrections and compare them with the other contributions to the
cross section. This can provide an estimate of the numerical
impact of the heavy flavor corrections once these are introduced
in a MC and correctly matched with the parton shower.
In Table~{\ref{tab1}} (above) we show the results of such an evaluation 
for $\sqrt{s} = 1$ GeV and  for
$50^\circ < \theta < 130^\circ$  (see \cite{hfbha} for details).
\begin{table*}
\begin{center}
{\rotatebox{90}{\makebox(0,0){\strut{} $\sqrt{s}=1$~GeV}}}
\hspace*{3mm}
\begin{tabular}{ c c c c c c c c }
\hline
$\theta$ & $phot$ ($10^{-4}$) & $e$ ($10^{-4}$) & $\mu$ ($10^{-4}$) & $c$ ($10^{-4}$)  &
$\tau$ ($10^{-4}$) & $b$ ($10^{-4}$) \cr
\hline
$50^{\circ}$  & 36.688225 & 17.341004 & 1.7972877 & 0.3605297 & 0.0264013 & 0.0040026 \cr
$70^{\circ}$  & 41.240039 & 19.438718 & 2.6504950 & 0.5795114 & 0.0465329 & 0.0065839 \cr
$90^{\circ}$  & 45.780639 & 21.463240 & 3.4581845 & 0.6528096 & 0.0576348 & 0.0077240 \cr
$110^{\circ}$ & 49.366078 & 23.099679 & 4.0922189 & 0.5082196 & 0.0495028 & 0.0065277 \cr
$130^{\circ}$ & 50.349342 & 23.847394 & 4.4392717 & 0.2421310 & 0.0273145 & 0.0039094 \cr
\hline
\end{tabular}
\end{center}
\begin{center}
{\rotatebox{90}{\makebox(0,0){\strut{} $\sqrt{s}=500$~GeV}}}
\hspace*{22mm}
\begin{tabular}{ c c c c c c c }
\hline
$\theta$ & $phot$ ($10^{-3}$) & $e$ ($10^{-3}$) & $\mu$ ($10^{-3}$) &
$\tau$ ($10^{-3}$) & $t$ ($10^{-3}$) \cr
\hline
$1^{\circ}$ & 7.0074592 & 3.4957072 & 0.9690710 & 0.1542329 & 0.0000575 \cr
$50^{\circ}$ & 14.819671 & 7.5740980 & 2.3185800 & 1.8411736 & 0.1707137 \cr
$70^{\circ}$ & 15.687591 & 8.0081541 & 2.3708714 & 2.0072240 & 0.2998535 \cr
$90^{\circ}$ & 16.560845 & 8.4172449 & 2.4207950 & 2.1521199 & 0.4202418 \cr
$110^{\circ}$ & 17.270026 & 8.7451035 & 2.4090920 & 2.2456055 & 0.4979010 \cr
$130^{\circ}$ & 17.512918 & 8.8954702 & 2.2543834 & 2.2446158 & 0.5287459 \cr
\hline
\end{tabular}
\caption{The second-order photonic, electron, muon,
$c$-quark, $\tau$-lepton, $b$-quark and $t$-quark contributions to the
differential cross section of Bhabha scattering at $\sqrt{s}=1$~GeV
and $\sqrt{s}=500$~GeV in units of $10^{-4}$ (above) and $10^{-3}$ (below)
of the Born cross section. The $c$, $b$ and $t$
contributions include the ${\mathcal O}(\alpha \alpha_s)$ term.
\label{tab1}}
\end{center}
\end{table*}
The muon loop diagrams dominate the heavy flavor corrections; they
are an order of magnitude larger than the corrections involving
heavier fermions. They reach 1/2 permille of the Born cross
section at large $\theta$.
In the lower part of Table~{\ref{tab1}}  we show the results
concerning the case in which $\sqrt{s} = 500$~GeV. In this case, the
muon and the tau contributions are of the same order, while the
top contribution is suppressed by an order of magnitude at
$90^{\circ}$ and by four orders of magnitude at $1^{\circ}$.

In conclusion, the calculation of the two-loop corrections to
Bhabha scattering in QED is now complete; these results remove the
last piece of pure theoretical uncertainty in the luminosity
determination at low- and high-energy accelerators.

\end{document}